\documentclass[%
 reprint,
 superscriptaddress,
 amsmath,amssymb,
 aps,
floatfix,
]{revtex4-2}

\usepackage[latin1]{inputenc}
\usepackage{bm}
\usepackage{multirow,amssymb,amsbsy,amsmath}
\usepackage{graphicx}
\usepackage{subfigure}
\usepackage{verbatim}
\usepackage{upgreek} %up体希腊字母
\usepackage[T1]{fontenc}
\makeatletter
\usepackage{pifont}
\usepackage{verbatim}
\makeatother

\begin{document}

% \preprint{APS/123-QED}

\title{Coherent Optical Memory Based on a Laser-Written On-Chip Waveguide}% Force line breaks with \\
% \thanks{A footnote to the article title}%

\author{Tian-Xiang Zhu}
\author{Chao Liu}
% \author{Tao Tu}
% \author{Zong-Feng Li}
\author{Liang Zheng}
\author{Zong-Quan Zhou}
\email{email:zq\_zhou@ustc.edu.cn}
\author{Chuan-Feng Li}
\email{email:cfli@ustc.edu.cn}
\author{Guang-Can Guo}

\affiliation{CAS Key Laboratory of Quantum Information, University of Science and Technology of China, Hefei 230026, China\\}
\affiliation{CAS Center For Excellence in Quantum Information and Quantum Physics, University of Science and Technology of China, Hefei 230026, China \\}
\date{\today }

\begin{abstract}
Quantum memory is the core device for the construction of large-scale quantum networks.
For scalable and convenient practical applications, integrated optical memories, especially on-chip optical memories, are crucial requirements because they can be easily integrated with other on-chip devices.
Here, we report the coherent optical memory based on a type-IV waveguide fabricated on the surface of a rare-earth ion-doped crystal (i.e. $\mathrm{Eu^{3+}}$:$\mathrm{Y_2SiO_5}$).
The properties of the optical transition ($\mathrm{{^7}F{_0}\rightarrow{^5}D{_0}}$) of the $\mathrm{Eu^{3+}}$ ions inside the surface waveguide are well preserved compared to those of the bulk crystal.
Spin-wave atomic frequency comb storage is demonstrated inside the type-IV waveguide.
The reliability of this device is confirmed by the high interference visibility of ${97\pm 1\%}$ between the retrieval pulse and the reference pulse.
The developed on-chip optical memory paves the way towards integrated quantum nodes.

\end{abstract}

%\keywords{Suggested keywords}%Use showkeys class option if keyword
                              %display desired
\maketitle

%\tableofcontents
\section{INTRODUCTION}

Optical quantum memories (QMs), as a light-matter interface, are enabling techniques for various applications in quantum information science, including the generation of on-demand photons \cite{lvovsky2009optical, nunn2008multimode}, serving as an identity quantum gate in quantum computation \cite{lvovsky2009optical, kok2007linear}, and building the quantum repeater for large-scale quantum networks \cite{briegel1998quantum, sangouard2011quantum}.
Among various physical systems for the implementation of QMs \cite{de2008solid, wang2019efficient, ritter2012elementary, chrapkiewicz2017high, wallucks2020quantum}, rare-earth-ion doped crystals, as a solid-state platform, have shown some unique advantages, such as large bandwidth \cite{usmani2010mapping, saglamyurek2011broadband, saglamyurek2016multiplexed, zhong2017interfacing}, high multimode capacity \cite{de2008solid, usmani2010mapping, sinclair2014spectral, tang2015storage, laplane2017multimode, kutluer2017solid, seri2017quantum, yang2018multiplexed, seri2019quantum}, long-lived coherence \cite{zhong2015optically, laplane2017multimode, ranvcic2018coherence}, and high storage fidelity \cite{de2008solid, zhou2012realization, jin2015telecom, zhou2015quantum, zhong2017interfacing}.
For scalable and convenient applications, great efforts have been devoted to the integrated optical memories based on waveguides manufactured by various techniques \cite{sinclair2010spectroscopic, saglamyurek2011broadband, thiel2012rare, liu2012femtosecond, sinclair2014spectral, jin2015telecom, saglamyurek2016multiplexed, corrielli2016integrated, seri2018laser,liu2020reliable, marzban2015observation, zhong2017nanophotonic}.
Femtosecond-laser micromachining (FLM) has been a powerful tool for the fabrication of integrated optical memory due to its advantages of high machining accuracy and low damages to the samples \cite{perry1999ultrashort, gattass2008femtosecond, skryabin2020femtosecond}.
Integrated optical memory \cite{corrielli2016integrated, liu2020reliable} and storage of single-photon-level light pulses \cite{seri2018laser, seri2019quantum} using FLM waveguides have been implemented with comparable performances in storage times and storage efficiencies as compared to those based on bulk materials.

Generally, there are four types of waveguide obtained by FLM \cite{chen2014optical}.
The previous demonstrations of integrated memories based on FLM focus on type-I \cite{seri2018laser, seri2019quantum} and type-II waveguides \cite{corrielli2016integrated, seri2018laser, liu2020reliable}.
These waveguides are fabricated at a depth of hundreds of micrometers beneath the surface of the crystal.
%However, the integrated optical devices should be made on the surface of the substrate to conveniently interface with other on-chip devices such as coplanar electrical structures.
While, integrated optical devices made on the surface of the substrate are more convenient to interface with other on-chip devices, such as coplanar electrical structures \cite{kindem2020control} and ultra-compact photonic structures \cite{turduev2018ultracompact}.

Here, we fabricate a type-IV on-chip waveguide in a $\rm Eu^{3+}$:$\rm{Y_2SiO_5}$ crystal and utilize it to conduct a coherent optical memory.
We carefully characterize the optical transition ($\rm{^7}F{_0}\rightarrow{^5}D{_0}$) of $\rm Eu^{3+}$ ions inside the type-IV waveguide, including the optical coherence time (${T_2}$) and the optical inhomogeneous broadening.
We implement the spin-wave atomic frequency comb (AFC) storage and observe high-visibility interference between the readout echo and a reference pulse.
The spin inhomogeneous broadening of $\rm Eu^{3+}$ ions inside the waveguide is characterized by analyzing the spin-wave storage dephasing of the AFC storage.

\section{FABRICATION OF THE TYPE-IV WAVEGUIDES}

\begin{figure*}[htbp]
 \centering
 \includegraphics[width=1\linewidth]{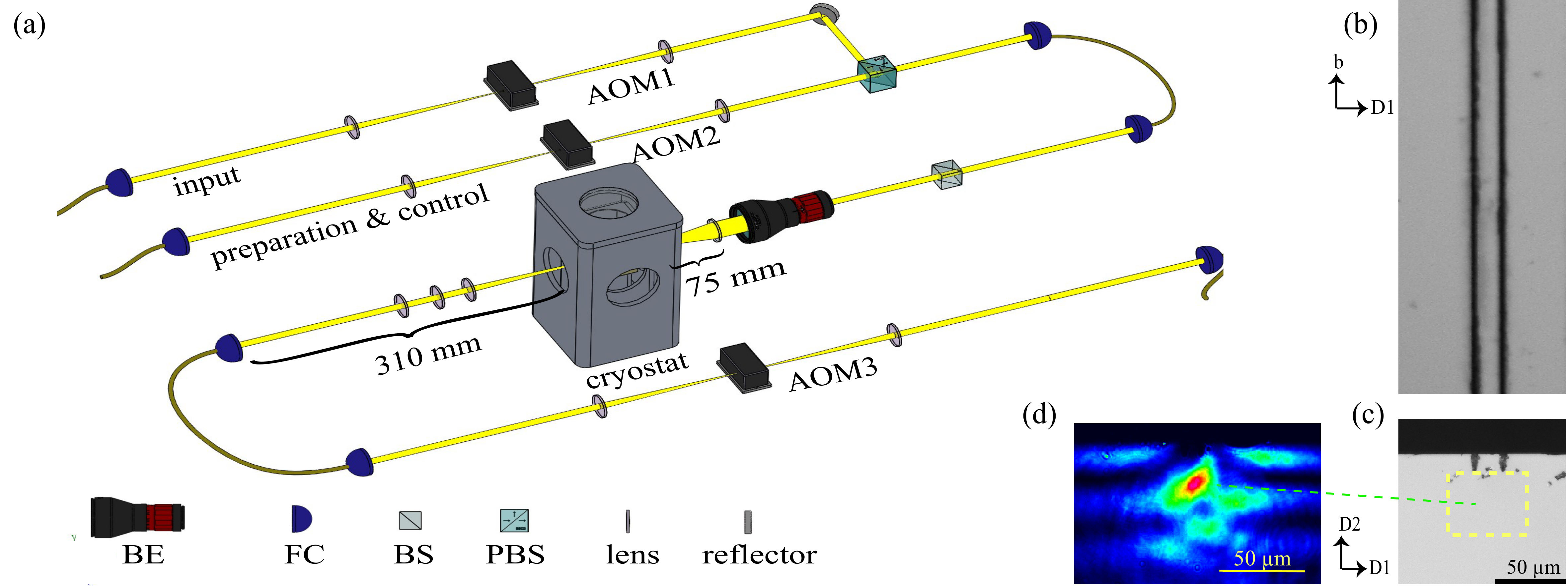}% Here is how to import EPS art  
 \caption{\label{setup} (color online)
 (a) Experimental setup. The crystal is placed in a cryostat with a base temperature of 3.2 K. The incident laser beam is split into two parts, i. e. the input mode and the preparation and control mode.
They are modulated by independent acousto-optic modulators (AOMs), and combined via a beam splitter (BS, with a reflectivity of 8\%) and finally coupled into the single-mode fiber. The Gaussian beam with a FWHM of 7 $\upmu$m is coupled into the type-IV waveguide via a 3$\rm\times$ beam expander (BE, GBE03-A, Thorlabs) and a lens with a focal length of 75 mm.
The output beam is coupled into the single-mode fiber by the lens set for spatial filtering.
Finally, the output beam is sent into a photoelectric detector after passing through a temporal gate based on AOM3. FC denotes fiber coupler and PBS denotes polarization beam splitter.
 (b) Top view of the type-IV waveguide under a microscope in the $b\times D1$ plane of the crystal. 
 (c) Front view of The type-IV waveguide under a microscope in the $D2\times D1$ plane of the crystal.
 The scale bar of (b) and (c) is 50.0 $\upmu$m.
  %The white area is crystal and the black area is air.
 The ridges have a width of 4.8 $\upmu$m and a depth of 13.0 $\upmu$m.
 (d) The guided mode intensity distribution at the exit surface of the crystal. The FWMH of the guided mode is about 33 $\upmu$m $\times$ 30 $\upmu$m (horizontal $\times$ vertical) at the exit surface of the crystal.
 The scale bar of (d) is 50.0 $\upmu$m.
}
\end{figure*}

The waveguides fabricated by FLM are classified according to the refractive index changes in the laser-written area, geometric shape and waveguide position of the sample \cite{chen2014optical}.
The type-IV waveguide consists of two deep laser-irradiated scores and is located on the surface of the crystal; it is also called as the ridge waveguide. 
The laser beam can be constrained between the two gaps and propagates through the waveguide.

We use the FLM system from WOPhotonics (Altechna RD Ltd, Lithuania) to fabricate the type-IV waveguide on a $\rm Eu^{3+}$:$\rm{Y_2SiO_5}$ crystal, which has a doping level of 0.1\% and a dimension of 10 mm$\times$ 3.25 mm$\times$3.25 mm ($b \times D1\times D2$).
Many parameters affect the results of the fabrication processes, for instance, the wavelength, the pulse duration, the polarization, the repetition frequency, the waist size of the focus spot, and the energy of per pulse of the femtosecond laser \cite{corrielli2016integrated, seri2018laser}.
Considering the anisotropy of the $\rm Eu^{3+}$:$\rm{Y_2SiO_5}$ crystal, the injection direction of the laser beam also influences the results of FLM \cite{seri2018laser, liu2020reliable}.
In our experiment, the femtosecond laser beam is focused on the surface of the crystal by a $20\times$ objective whose numerical aperture is 0.4.
The laser beam injects along the $D2$ axis of the crystal and moves along the b axis with a speed of 1 mm/s.
The distance between the two parallel damage tracks is set as 20 $\upmu$m.
By optimizing the single-mode coupling efficiency of the 580-nm laser polarized along the $D1$ axis in type-IV waveguides fabricated with various conditions, we obtain the following parameters: 1030 nm laser wavelength with 300 fs pulse duration, 20 kHz repetition rate, 1.7 $\upmu$J per-pulse energy and the polarization of the laser along the $b$ axis.
The guided mode is bound by two ridges and the upper surface of the crystal.
Since the defects on the crystal also affect the performance of FLM \cite{liu2020reliable}, we fabricate several type-IV waveguides in the $\rm Eu^{3+}$:$\rm{Y_2SiO_5}$ crystal, in order to find a high-quality waveguide to conduct the experiments. 
The direct transmission (include coupling mode mismatch, waveguide propagation loss and Fresnel losses) of the waveguide is 19.4\%.
The coupling efficiency of the waveguide output into the single-mode fiber is 54\%.
The full width at half maximum (FWHM) of the guided mode is about 33 $\upmu$m $\times$ 30 $\upmu$m (horizontal $\times$ vertical) at the exit surface of the crystal, as shown in Fig. 1(d). 

\begin{figure*}[htbp]
 \centering
 \includegraphics[width=1\linewidth]{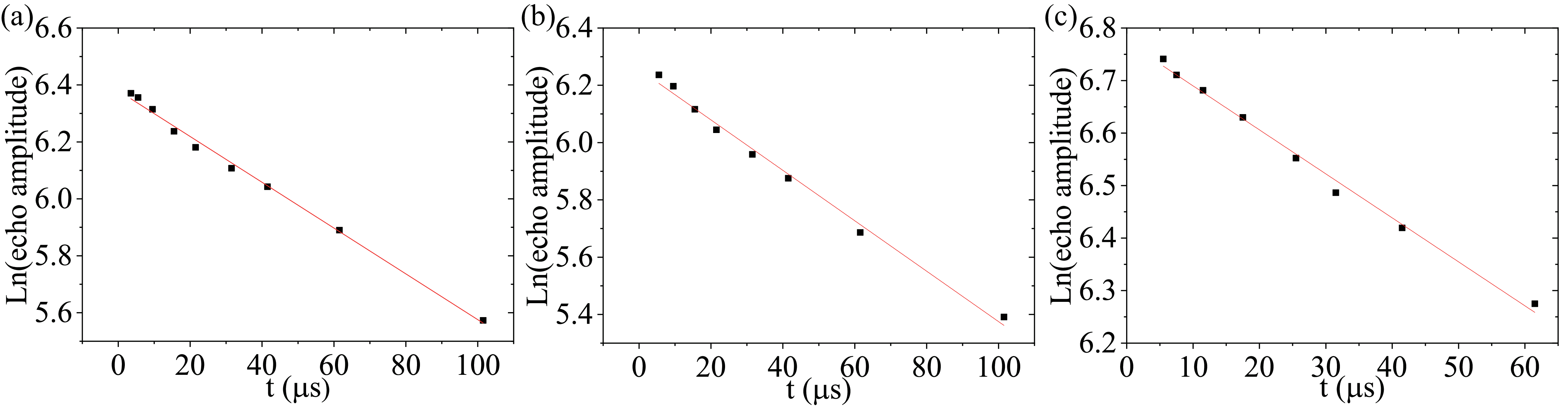}
 \caption{\label{T2type4} (color online)
 Two-pulse photon echo measurements implemented in three different regions inside the sample where $t$ is the spacing between the two incident pulses.
 The red solid lines are the results of the linear fitting.
 (a) In the type-IV waveguide, the coherence  time ${T_2}$ is fitted as $124\pm4$ $\upmu$s.
(b) In the bulk-1 area, ${T_2}$ is fitted as $113\pm3$ $\upmu$s.
(c) In the bulk-2 area, ${T_2}$ is fitted as $119\pm4$ $\upmu$s.
}
\end{figure*}

\section{EXPERIMENT SETUP}

The laser source is a frequency-doubled semiconductor laser at 580 nm (TA-SHG, Toptica).
The laser is locked to a stabilized Fabry-Perot (FP) cavity using the Pound-Drever-Hall technique \cite{black2001introduction}.
The FP cavity is constructed from ultra-low expansion glasses and has a linewidth of 50 kHz. The locked laser has a linewidth of the order of 0.1 kHz.
We use the AOMs and an arbitrary waveform generator to modulate and control the laser beam.
As shown in Fig.~\ref{setup}, AOM1 and AOM2 are used to modulate the input mode and preparation and control mode, respectively.
The sample is assembled on a cryostat (Montana Instruments) that can provide an environment of low vibration and a base temperature of approximately 3.2 K.
Besides, a piezo-actuated three-axis stage is installed on the cryostat and its accuracy is of the order of 10 nm, which makes it convenient to couple the laser beam into the waveguides. 

\section{RESULT}
\subsection{Characterization of the type-IV waveguide}

To check whether the optical properties of the type-IV waveguide are modified as compared to those of the bulk area, we first measure the optical inhomogeneous broadening of the optical transition ($\rm{^7}F{_0}\rightarrow{^5}D{_0}$) of $\rm Eu^{3+}$ ions in different sections of the crystal.
We choose three representative positions of the crystal, including the type-IV waveguide region, the bulk section at the same depth (13 $\upmu$m) beneath the upper surface of the crystal as the waveguide region (marked as bulk 1), and the bulk section at the center region, which is 1.6 mm beneath the top surface of the crystal (marked as bulk 2).
We scan the frequency of incident light to obtain the absorption spectrum of the sample.
The measured peak absorption frequency of bulk 1 is $516\thinspace847.6\pm0.1$ GHz and its FWHM is $2.0\pm0.2$ GHz; the peak absorption frequency of bulk 2 is $516\thinspace847.6\pm0.1$ GHz and its FWHM is $2.0\pm0.2$ GHz.
These results indicate that the near-surface section of the crystal has the same optical inhomogeneous broadening as the center section of the crystal.
The peak absorption frequency of the waveguide region is $516\thinspace847.2\pm0.1$ GHz and its FWHM length is $2.5\pm0.1$ GHz.
The maximum absorption frequency of the waveguide has a red shift of 400 MHz compared to that of the two bulk sections and the linewidth is slightly broadened by 500 MHz.

The optical coherence time ($T_2$) is another important parameter to characterize the coherent properties of the type-IV waveguide.
Two-pulse photon echo \cite{tittel2010photon} experiments are implemented in the three representative regions (the type-IV waveguide area, bulk-1 area and bulk-2 area) of the sample.
In Fig.~\ref{T2type4} we shows the two-pulse photon echo amplitude as a function of the two-pulse spacing; the optical coherence time ($T_2$) is extracted from the inverse of the slope of the fitted curve.
In the type-IV waveguide, we get $T_2$ of $124\pm4$ $\upmu$s when the peak power of the input pulse is 1.8 mW.
The $T_2$ is $113\pm3$ $\upmu$s in the bulk 1 area and $119\pm4$ $\upmu$s in the bulk 2 area, with the peak power of the input pulse 4.7 and 4.6 mW, respectively.
When testing in the bulk-1 and the bulk-2 areas, the input mode is the same as the optical mode coupling into the waveguide and has a FWHM of 7 $\upmu$m in the input surface of the crystal.
The obtained $T_2$ for the waveguide region is essentially the same as that measured in the bulk regions. Therefore, we conclude that the fabrication process of the type-IV waveguide has negligible effects on the optical coherent properties of the sample, and this on-chip waveguide is ready for use as an optical quantum memory.
The $T_2$ value reported here is partially limited by the ``instantaneous spectral diffusion'' because of the nonvanishing excitation \cite{konz2003temperature}. Furthermore, the sample temperature may be slightly higher than expected since the sample is loosely connected to the cold finger to avoid any damage to the surface waveguide.

\subsection{Optical storage based on the type-IV waveguide}

\begin{figure}[htbp]
 \centering
 \includegraphics[width=1\linewidth]{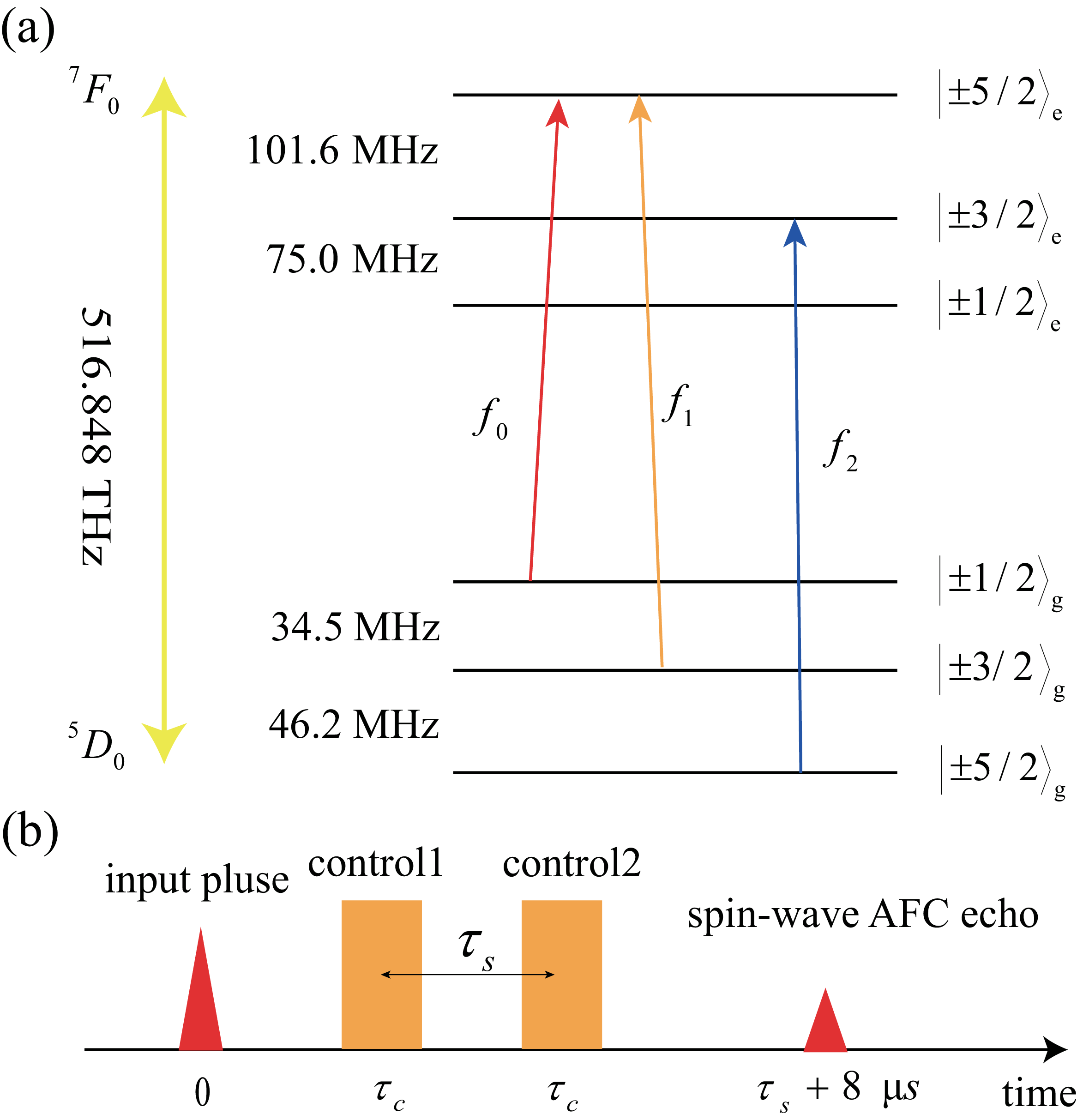}% Here is how to import EPS art  
 \caption{\label{levelandprocess} (color online)
(a)	The hyperfine energy level diagram of the $\rm{^7}F{_0}\rightarrow{^5}D{_0}$ transition of  
$\rm^{151}Eu^{3+}$ ions in the $\rm{Y_2SiO_5}$ crystal at zero magnetic field.
The input signal is resonant with the $|\pm1/2\rangle_g\rightarrow|\pm5/2\rangle_e$ transition with a frequency of $f_0$.
The control pulse is resonant with the $|\pm3/2\rangle_g\rightarrow|\pm5/2\rangle_e$ transition with a frequency of $f_1$.
The $|\pm5/2\rangle_g\rightarrow|\pm3/2\rangle_e$ transition with a frequency of $f_2$ is necessary for the AFC preparation process, which helps to isolate a single class of ions \cite{liu2020reliable}.
(b) The time sequence for the spin-wave AFC scheme.
The duration of the control pulse (${\tau_c}$) is 2.5 $\upmu$s.
}
\end{figure}

The AFC scheme is a widely employed quantum memory scheme in the solid-state system with the advantages of large bandwidth \cite{usmani2010mapping, saglamyurek2011broadband, saglamyurek2016multiplexed, zhong2017interfacing}, high multimode capacity \cite{de2008solid, usmani2010mapping, sinclair2014spectral, tang2015storage, seri2017quantum, yang2018multiplexed, seri2019quantum}, long-lived coherence \cite{zhong2015optically, laplane2017multimode, ranvcic2018coherence}, and high storage fidelity \cite{de2008solid, zhou2012realization, jin2015telecom, zhou2015quantum, zhong2017interfacing}.
Taking advantage of the large inhomogeneous broadening of the optical transition ($\rm{^7}F{_0}\rightarrow{^5}D{_0}$) of $\rm Eu^{3+}$ ions, a comb structure can be shaped on the spectrum by utilizing a spectral-hole burning technique \cite{timoney2013single}.
The periodicity of the comb is denoted by $\Delta$.
When an optical pulse is absorbed by $\rm Eu^{3+}$ ions in the AFC at the initial moment, those ions with a detuning of $m\Delta$ will be collectively excited where $m$ is an integer.
During the dephasing of the excited ions, a coherent AFC photon echo emits at time $1/\Delta$ \cite{afzelius2009multimode}.
Long storage time and on-demand retrieval can be achieved by further utilizing the spin-wave AFC scheme \cite{timoney2013single}.
The core of this method is utilizing a control pulse to transfer the ions from the excited state into the extra empty ground state that freezes the AFC evolution and suppresses the two-level AFC echo.
Then, another control pulse is applied to bring back the excitation.
The rephasing process of the AFC scheme will continue and emit an echo (the so-called spin-wave AFC echo). The total storage time of the spin-wave AFC memory is ${t_{tot}=\tau_s+1/\Delta}$ with $\tau_s$ the time spacing between the two control pulses.

\begin{figure}[htbp]
 \centering
 \includegraphics[width=1\linewidth]{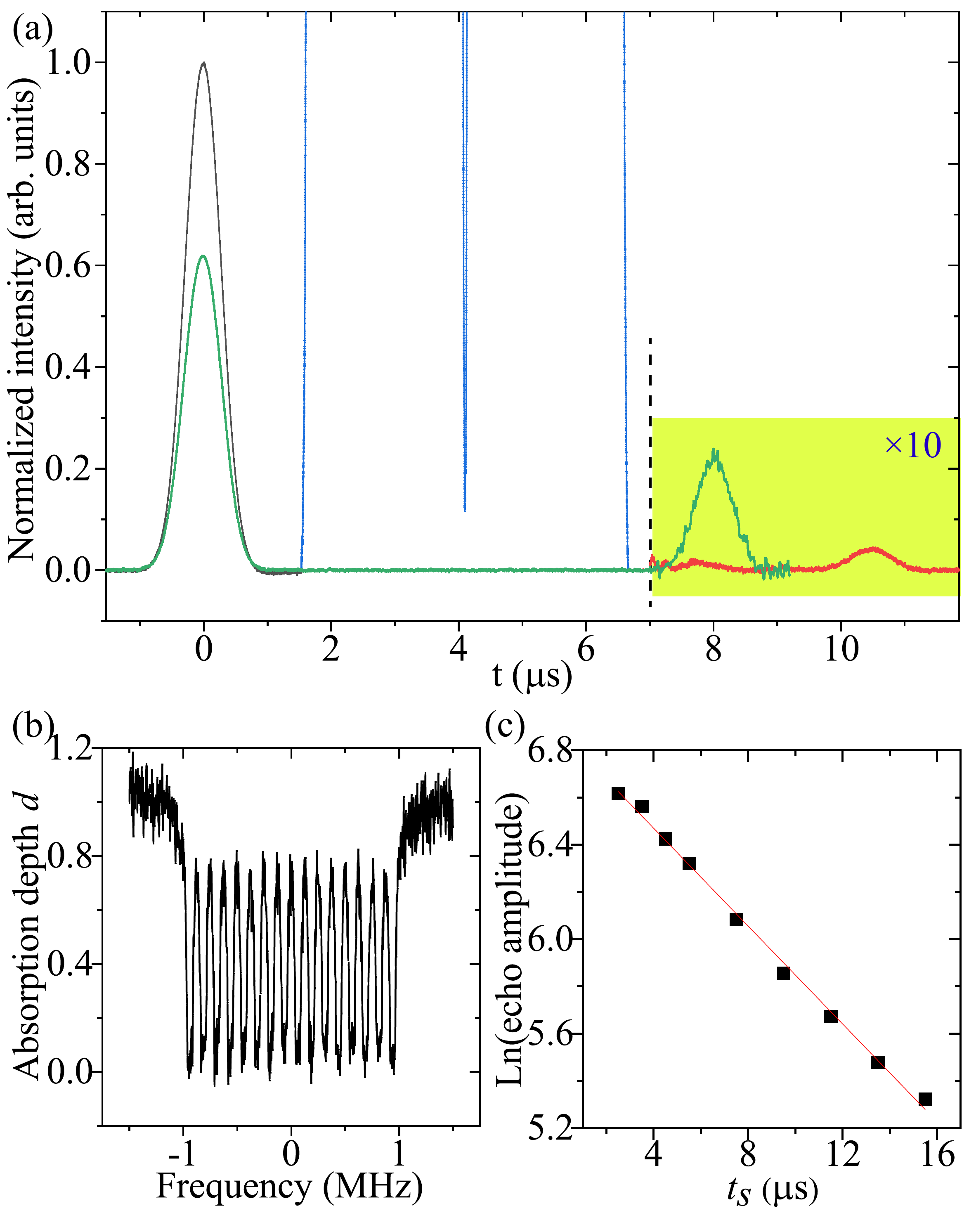}% Here is how to import EPS art  
 \caption{\label{spinlinewidth} (color online)
(a) The temporal traces of the input pulse (black line), the AFC echo (green line) and the spin-wave AFC echo (red line). The blue line indicates the positions of the two control pulses and the traces in the yellow region are magnified 10 times.
(b) The prepared AFC at the $|\pm1/2\rangle_g \rightarrow |\pm5/2\rangle_e$ transition. The peak comb absorption depth ($d$) is $0.79\pm0.02$, the background absorption depth ($d_0$) is  $0.00\pm0.03$ and the finesse ($F$) of the combs is about $2.6\pm0.3$.
(c) Exponential decay of the spin-wave AFC echo amplitude with storage time in the spin state ($t_s$).
The red solid line is the result of the linear fitting, and the inhomogeneous linewidth of the spin transition ($|\pm1/2\rangle_g \rightarrow |\pm3/2\rangle_g$) is $33\pm1$ kHz.
}
\end{figure}

The $\rm{^7}F{_0}\rightarrow{^5}D{_0}$ level structure of the $\rm^{151}Eu^{3+}$ ions in the $\rm{Y_2SiO_5}$ crystal is shown in Fig.~\ref{levelandprocess}.
We utilize the spectral hole-burning process \cite{liu2020reliable, lauritzen2012spectroscopic, jobez2016towards} to prepare an AFC comb structure inside the waveguide, including class cleaning, population polarization, and creating the AFC structure at $|\pm1/2\rangle_g$ while keeping $|\pm3/2\rangle_g$ empty.
The details of the preparation process can be found in our recent work \cite{liu2020reliable}.
As shown in Fig.~\ref{spinlinewidth}, the prepared AFC has a bandwidth of 2 MHz and the measured internal efficiency of the AFC echo is 2.4\% at a storage time of 8 $\upmu$s, which is calculated as the ratio between the intensity of the AFC echo and the input pulse.
The theoretical AFC storage efficiency \cite{de2008solid} as estimated from the AFC structure shown in Fig. 4(b) is 2.4\% which agrees with the measured value.
The storage efficiency is primarily limited by the low absorption of the sample since the dopant has two natural isotopes ($\rm ^{151}Eu$ and $\rm^{153}Eu$) and only $\rm^{151}Eu^{3+}$ is employed here.
At time $t=0$, an input pulse with a frequency of $f_0$ and FWHM of 750 ns is injected into the type-IV waveguide and its AFC echo appears after 8 $\upmu$s.
Next, two control pulses with a center frequency of $f_1$ and a chirp bandwidth of 2 MHz are applied to implement the spin-wave AFC scheme.
The peak power of the control pulses is 58 mW before the cryostat and they each have a duration ($\tau_c$) of 2.5 $\upmu$s. 
The required power of the control pulses is approximately 10 times lower than that required in the bulk material \cite{jobez2015coherent}.
The estimated transfer efficiency of a single control pulse is approximately 50\%. 
The spin-wave echo appears at $t_{\mathrm{tot}}=\tau_s+8$ $\upmu$s.
The peak power of the input pulse is 5.8 mW before the cryostat in spin-wave storage and the internal efficiency of the 10.5 $\upmu$s spin-wave AFC storage is 0.5\% inside the waveguide.

Because of the spin dephasing, the spin-wave AFC echo decays when $\tau_s$ increases.
As shown in Fig.~\ref{spinlinewidth}(c), the spin-wave AFC echo amplitude can be linear fitted with storage time in the spin state ($\tau_s$) on a log scale, which indicates a Lorentzian profile of the spin inhomogeneous broadening for $|\pm1/2\rangle_g \rightarrow |\pm3/2\rangle_g$ \cite{beavan2012photon, serrano2018all}.
The echo amplitude ($A$) can be represented by
\begin{equation}
 {A}=A(0)e^{-\pi \gamma_s \tau_s}
\label{eq1}
\end{equation}
where $\gamma_s$ is the inhomogeneous linewidth of the spin transition and $A(0)$ is the spin-wave AFC echo amplitude with $\tau_s=0$.
The $\gamma_s$ is fitted as $33\pm1$ kHz in our type-IV waveguide.
The spin inhomogeneous broadening of $\rm Eu^{3+}$ ions inside the type-IV waveguide is approximately unchanged as compared to what is usually measured in the bulk region of a similar crystal \cite{jobez2015coherent}, showing that the spin transition is also well protected in the waveguide fabrication process.

\begin{figure}[htbp]
 \centering
 \includegraphics[width=1\linewidth]{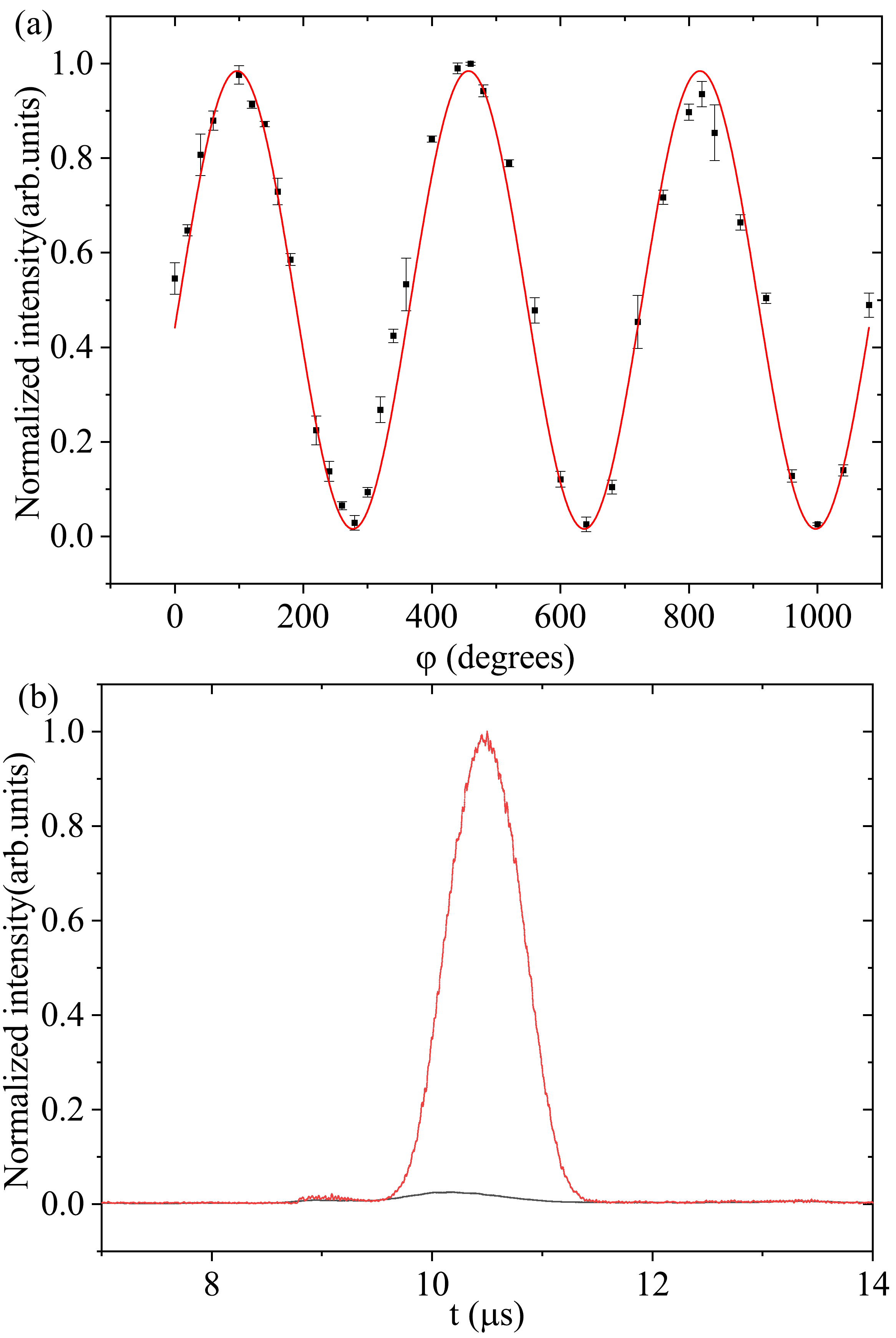}% Here is how to import EPS art  
 \caption{\label{interference} (color online)
 (a) Interference between the readout echo and a reference pulse. Here $\phi$ corresponds to the relative phase between the input pulse and the reference pulse.
 The data points are fitted using ${I(\varphi)=(I_{max}/2)[1+V \sin (\varphi+\varphi')]}$ \cite{lovric2013faithful, liu2020reliable}, where ${I_{max}}$ is the maximum intensity, $V$ is the interference visibility, and ${\varphi'}$ is the initial phase. The fitting curve is represented by the red line with \(V=0.97\pm0.01\).
 (b) The traces of the readout echo in the cases of constructive (red line) and destructive (black line) interference.
 }
\end{figure}

To characterize the fidelity of the entire storage process, we implement an interference test between the spin-wave AFC echo and a coherent reference pulse.
The reference pulse is tuned precisely to overlap with the readout echo in the time domain \cite{lovric2013faithful, liu2020reliable}.
Meanwhile, the relative phase \(\varphi\) between the signal pulse and the reference pulse is modulated to obtain the interference curve.
The interference between them is detected using a calibrated photomultiplier tube \cite{liu2020reliable}.
We used the same AFC preparation sequence, input pulse and control pulse as the previous experiment.
We set \(t_s=2.5\) $\upmu$s and record the signal when increasing the relative phase \(\varphi\) with a step of 30 degrees.
The results are shown in Fig.~\ref{interference} with a fitted interference visibility $V$ of $0.97\pm0.01$.
This high visibility demonstrates the reliability of the on-chip optical memory in coherent phase storage.
The remaining imperfection is mainly attributed to the imperfection of waveform overlap between the signal and reference pulses and the detector noise.

\section{CONCLUSION}

A laser-written on-chip type-IV waveguide is fabricated on a $\rm Eu^{3+}$:$\rm{Y_2SiO_5}$ crystal.
Coherent optical storage based on the spin-wave AFC scheme is implemented with high fidelity. 
The fabrication of the type-IV waveguide slightly shifts the absorption center and expands the optical inhomogeneous broadening, while the optical coherence time is unchanged during the fabrication.
The spin inhomogeneous broadening of $\rm Eu^{3+}$ ions inside the type-IV waveguide is approximately unchanged as compared to what is usually measured in the bulk region of a similar crystal \cite{jobez2015coherent}.
This is in contrast to the type-II waveguide fabricated in $\rm Eu^{3+}$:$\rm{Y_2SiO_5}$ that introduces a significant broadening of both the optical transition and the spin transition \cite{liu2020reliable}.
This can be explained by the fact that the fabrication process of the type-IV waveguide is less damaging to the crystal than that of the type-II waveguide.

Because of its on-chip structure, the type-IV waveguide has unique benefits for future practical applications.
Nevertheless, several extensions of the current work can be considered.
First, the coupling efficiency is currently limited by the rough edges of the ridges, which can be improved by using a higher magnification objective to increase the accuracy of FLM, constructing the new ridges by multiple micromachinings at different depths under low per-pulse energy, or using ion-beam sputtering to reduce the roughness of the edges \cite{chen2014optical}.
Second, the spin-wave AFC scheme can be combined with dynamic decoupling to extend the storage lifetime \cite{fraval2005dynamic, jobez2015coherent}.
By combining the type-IV waveguide with coplanar electrical waveguides, it is possible to achieve high-efficiency dynamical decoupling control with reduced radio-frequency power requirements. The efficient interface between such on-chip photonic memory and other on-chip photonic circuits should lead to versatile applications in large-scale quantum networks. 

\section{acknowledgments}

This work is supported by the National Key R\&D Program of China (Grant No. 2017YFA0304100), the National Natural Science Foundation of China (Grants No. 11774331, 11774335, No. 11504362, No. 11821404, and No. 11654002), the Anhui Initiative in Quantum Information Technologies (Grant No. AHY020100), the Key Research Program of Frontier Sciences, CAS (Grant No. QYZDY-SSW-SLH003), the Science Foundation of the CAS (Grant No. ZDRW-XH-2019-1), the Fundamental Research Funds for the Central Universities (Grants No. WK2470000026 and No. WK2470000029), and the Youth Innovation Promotion Association CAS.
% % The \nocite command causes all entries in a bibliography to be printed out
% % whether or not they are actually referenced in the text. This is appropriate
% % for the sample file to show the different styles of references, but authors
% % most likely will not want to use it.
% \nocite{*}
%\nocite{*}
%\bibliographystyle{apsrev4-2}
\bibliography{Coherent_Optical_Memory_Based_on_A_Laser-written_On-chip_Waveguide}% Produces the bibliography via BibTeX.

\end{document}